\documentclass[conference]{IEEEtran}
\usepackage{cite}
\usepackage{amsmath,amssymb,amsfonts}
\usepackage{algorithmic}
\ifCLASSINFOpdf
   \usepackage[pdftex]{graphicx}
\else
   \usepackage[dvips]{graphicx}
\fi
\usepackage{textcomp}
\usepackage{xcolor}
\usepackage[caption=false,font=footnotesize]{subfig}
\usepackage{booktabs} 
\usepackage{stfloats}
\usepackage{balance}
\usepackage{soul}
\usepackage[normalem]{ulem}
\usepackage{cancel}
\usepackage{verbatim}
\usepackage{url}
\usepackage{cases}

\begin{document}

\title{Augmenting Cloud Connectivity with Opportunistic Networks for Rural Remote Patient Monitoring 
\\
}

\author{
    \IEEEauthorblockN{Esther Max-Onakpoya\IEEEauthorrefmark{1}, Oluwashina Madamori\IEEEauthorrefmark{1}, Faren Grant\IEEEauthorrefmark{5}, Robin Vanderpool\IEEEauthorrefmark{2}\IEEEauthorrefmark{3}, Ming-Yuan Chih\IEEEauthorrefmark{4}\IEEEauthorrefmark{3},\\ David K. Ahern\IEEEauthorrefmark{7}\IEEEauthorrefmark{8}, Eliah Aronoff-Spencer\IEEEauthorrefmark{6}\IEEEauthorrefmark{5}, Corey E. Baker\IEEEauthorrefmark{1}}
    \IEEEauthorblockA{
        \IEEEauthorrefmark{1}Department of Computer Science, University of Kentucky, Lexington, KY, USA \\
    }
    \IEEEauthorblockA{
        \IEEEauthorrefmark{2}Department of Health, Behavior \& Society, University of Kentucky, Lexington, KY, USA \\
    }
     \IEEEauthorblockA{
        \IEEEauthorrefmark{3}Markey Cancer Center, University of Kentucky, Lexington, KY, USA\\
    }
    \IEEEauthorblockA{
        \IEEEauthorrefmark{4}Department of Health and Clinical Sciences, University of Kentucky, Lexington, KY, USA \\
    }
    \IEEEauthorblockA{
        \IEEEauthorrefmark{5}Qualcomm Institute, University of California San Diego, La Jolla, CA, USA \\
    }
    \IEEEauthorblockA{
        \IEEEauthorrefmark{6}Division of Infectious Diseases, University of California San Diego Health-Jacobs Medical Center, La Jolla, CA, USA \\
    }
    \IEEEauthorblockA{
        \IEEEauthorrefmark{7}Federal Communications Commission, Washington DC, USA \\
    }
    \IEEEauthorblockA{
        \IEEEauthorrefmark{8}Digital Behavioral Health and Informatics, Brigham and Women's Hospital, Harvard Medical School, Boston, MA, USA \\
        \{Esther.Max05, shina\}@uky.edu, fjgrant@ucsd.edu, robin@kcr.uky.edu, m.chih@uky.edu, \\ dahern@bwh.harvard.edu, earonoffspencer@ucsd.edu, baker@cs.uky.edu
    }
}

\maketitle

\begin{abstract}
Current remote patient monitoring (RPM) systems are fully reliant on the Internet. 
However, complete reliance on Internet connectivity is impractical 
in rural and remote environments where modern 
infrastructure is often lacking, power outages are 
frequent, and/or network connectivity is sparse (e.g. 
rural communities, mountainous regions of Appalachia, American Indian reservations, developing countries, and natural disaster situations). 
This paper proposes augmenting intermittent Internet with opportunistic communication 
to leverage the social behaviors of patients, caregivers, 
and community members to facilitate out-of-range monitoring of patients via Bluetooth 5 
during intermittent network connectivity in rural communities. 
The architecture is evaluated for Owingsville, KY using U.S. Census Bureau, the National Cancer Institute's, and IPUMS ATUS sample 
data, and is compared against a delay tolerant RPM case that is 
completely disconnected from the Internet.
The findings show that with only 0.30 rural adult population participation, 
the architecture can deliver 0.95 of non-emergency medical information with an average delivery latency of $\sim$13 hours. 
\end{abstract}


\section{Introduction}\label{sec_intro}
%
The ubiquity of mobile devices and rapid improvement in wireless body sensors has revolutionized the field of healthcare.
Through mHealth solutions, practitioners can remotely monitor and assist with patients' disease management in real time or asynchronously.
This has improved the timeliness of clinical decision making, decreased the length of hospital stays, and reduced mortality rates~\cite{niksch2014value,moy2017leading}. 
Although many patients have benefited from mHealth solutions, and national efforts are underway to accelerate broadband deployment in under-served areas of the US~\cite{usda2019}, rural patients may not benefit to the same extent as their non-rural counterparts due to geographical and financial barriers that result in limited or nonexistent access to broadband connectivity~\cite{henley2017invasive}.
Additionally, chronic disease is approximately 20\% more prevalent in rural areas than other areas~\cite{services2017}.

A major limitation of mHealth solutions in rural areas is financial burden. 
The cost associated with deploying infrastructure that facilitates complete connectivity is often too expensive for many rural cities to afford without significant financial assistance~\cite{usda2019}. Likewise, cost remains a barrier to adoption even when broadband may be accessible.
For example, the KentuckyWired project in Appalachian Kentucky costs at least \$324 million, with taxpayers having to pay \$1.5 billion over 30 
years\footnote{Lexington Herald Leader, ``Cost overruns in troubled Kentucky broadband project near 
\$100 million, audit finds'', 9/27/18, \url{https://www.kentucky.com/latest-news/article219058225.html}}.
A promising solution lies in the use of a network architecture that leverages human mobility for disseminating patient health information (PHI).

Though there has been some important research conducted to connect rural areas 
using delay tolerant networks (DTNs)~\cite{pentland2004daknet,doria2002providing,whitbeck2010hymad,raffelsberger2013hybrid}, the network characteristics
such as node mobility and density differ vastly resulting in the need for a better 
understanding of how DTNs will operate in rural remote patient monitoring (RRPM) scenarios.
A survey conducted in~\cite{syed2012study} that consisted of clinical and non-clinical staff 
showed that DTNs are believed to be a promising solution for remote patient monitoring in 
low-resource settings for number of application domains including EHRs, notifications, 
news, blogs, etc.
Additionally, Syed et. al. also report that there has been 
limited work on evaluating how DTNs will perform in RRPM environments~\cite{syed2012study}.
The works that have explored DTNS for RRPMs have investigated the use of vehicles to deliver patient health information (PHI)~\cite{barua2014rcare}
or using ad hoc networks within a hospital~\cite{cho2008opportunistic}.
Understanding how the use of human mobility influences data dissemination 
in RRPM environments is essential, but can be cumbersome due to its inherent characteristics.
Unlike other networks, such as social networks, in which most of the 
population is able to actively participate; patient monitoring networks typically consist of a small 
percentage of the population: patients, caregivers, and healthcare 
providers. 

Furthermore, RPM systems consist of a vast diversity of domain applications 
with different network performance requirements. 
For example, a pacemaker monitoring application has different network requirements from a distress measurement application. 
Therefore, it is imperative to understand the limitations posed by the domain application in order to design the right network for it.
Hence, we propose a novel hybrid architecture that leverages minimal Internet 
infrastructure along with node mobility for the dispersal of non-emergency PHI for RRPM within a community.

The contributions of this work are: (1) defines a hybrid architecture consisting of intermittent D2D and Internet connectivity for RRPM; 
(2) derives a simple and novel mathematical description for remote monitoring in rural communities which can be used by decision makers to design an optimal network topology that leverages the available infrastructure;
(3) provides insights on how the intrinsic characteristics of a real rural city, Owingsville, KY,
influences network performance of rural patient data when the dissemination of 
information depends solely on inherent community participation. 

The remainder of the paper is structured as follows:
Section~\ref{section:related} discusses the related work;
Section~\ref{section:hybrid_model} describes the network and its key members; 
Section~\ref{section:real_world} introduces the mathematical model;
Section~\ref{section:evaluation} offers the results obtained from the model simulation; 
Section~\ref{section:discussion} discusses the limitations of the model along with future work;
and finally, Section~\ref{section:conclussion} concludes the paper.

\section{Related Work}\label{section:related}
With regards to RRPM, some mobile ad-hoc network schemes have been proposed for 
transmitting PHI between patients and medical staff within a hospital~\cite{rashid2005real,cho2008opportunistic}.
Other researchers have investigated propagating voicemails or other emergency 
data via DTNs amongst communities~\cite{adlassnig2009rural,jang2009rescue}. 
More closely related work proposes using vehicular ad hoc networks to disseminate 
PHI in rural communities~\cite{barua2014rcare,murillo2011application}.
A major limitation of the aforementioned work is the lack of proof that 
evaluations adequately reflect real rural environments, particularly when it 
comes to node placement, mobility, density, and Internet connectivity; which 
are all required to properly assess solutions in rural environments. 

Additionally, the aforementioned work solely utilizes vehicular communications 
in transmitting messages to medical entities.
In regions where vehicular networks (such as bus systems, etc) do not exist, 
the DTN would have to depend on other forms of node mobility.
Hence, the use of DTN for RRPM requires an analysis of increasing user 
participation within the current density constraints of the respective community.
This is important because rural communities do not have many new people 
moving into or passing through their neighborhoods. 
Simply adding nodes to increase density in a simulation may not reflect the 
true behavior of how participating nodes will behave in the network.

Unlike other works, we propose a hybrid architecture that is able to facilitate 
RRPM by functioning as a DTN when Internet availability is completely non-existent,
and simultaneously leverage Internet connectivity when it intermittently becomes available. 
Next, we analyze the key factors that influence the viability 
of the network design for the attributes of a real rural city, Owingsville, KY.
Lastly we evaluate how our proposed system can opportunistically leverage 
the natural mobility of nodes, intrinsic density of a rural city, and inherent 
intermittent cloud connectivity for the RRPM problem.

\section{Hybrid Network Model}\label{section:hybrid_model}
The proposed hybrid network consists of a DTN that supplements an intermittent Internet-connected network.
The DTN consists of device-to-device (D2D) communication 
in order to facilitate out-of-range monitoring of patients. 
Using a hybrid network, the system is able to harness the mobility of members of the rural population and maximize data delivery without exceeding an understood latency.
Although this work solely focuses on a RRPM network design, the model described can be extended to represent any other hybrid network, including heterogeneous inter-band networks and intra-band networks.
The specifics of the evaluated hybrid network are discussed in Section \ref{sec_hybrid_arch_desc}.

\subsection{Nodes and Network Entities}\label{subsection:rrpm}

\subsubsection{Patient}\label{subsubsection:patient}
The primary data generator in a RRPM system is the patient. 
Objective patient data can be obtained from biosensors/wearables that form a body sensor network and transmitted asynchronously from a patient to a medical entity.
Conversely, subjective data, such as surveys and assessment forms can also be collected and transmitted to the patient's corresponding medical entity.

\subsubsection{Caregiver}\label{subsubsection:caregiver}
In managing chronic illnesses, most patients have at least one caregiver that actively participates in the patient's care~\cite{hughes2008patient}. 
Unlike the patient, they are often more mobile yet they remain close enough to the patient to attend to their needs. 
Hence, the proposed model classifies caregivers as active transport agents because they are the nodes that have both high access to patients and high mobility.
Additionally, caregivers are likely to encounter points of interests (POIs) and intermediary nodes when running errands for the patient.

\begin{figure*}[htbp]
	\centering
	\includegraphics[width=\textwidth, height=3.65in]{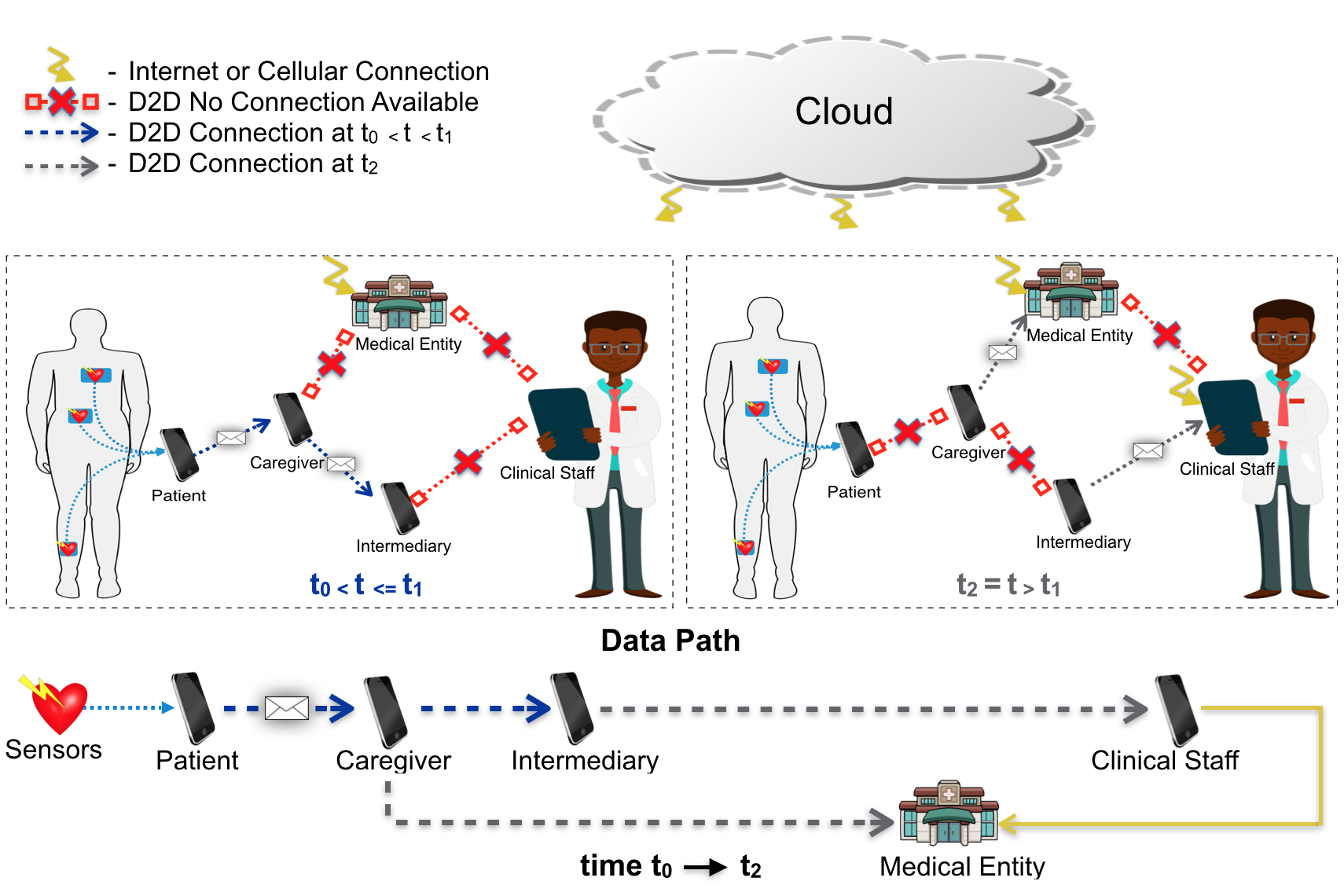}
	\caption{Hybrid Network for Rural Remote Patient Monitoring}
	\label{fig_opportunistic_path}
\end{figure*}

\subsubsection{Intermediary Network}\label{subsubsection:intermediary}
The intermediary network consists of nodes that are members of the society and already participate in other networks such as mail-delivery systems, trash collection, school bus transportation systems, and the like. 
The intermediary network can be leveraged to opportunistically collect and deliver data without significantly increasing the costs of deploying Internet in rural areas.
In relation to the RRPM network, they are classified as 
primary transport agents that frequently encounter POIs and caregiver nodes.
Intermediary nodes move through the network differently depending on their employment state and they are more likely to have Internet connectivity at some point during a 24 hour period. 

\subsubsection{Clinical Staff}\label{subsubsection:clinicalstaff}
The clinical staff network are nodes whose primary employment location is with the medical entity. 
These nodes are mobile and have the highest probability of encountering the destination. 

\subsubsection{Points of Interests}\label{subsubsection:pois}
POIs are stationary nodes or locations where transport agents typically congregate. 
Some examples include grocery markets, post offices, banks, places of worship, places of work, and city halls.
Internet access points can be added to POIs based on budget constraints.

\subsubsection{Destination}\label{subsubsection:destination}
The patient's associated medical entity is the primary destination in this model. 
Unlike other nodes, the medical entity is stationary and is less likely to have random encounters with the intermediary network.
Nevertheless, it is fully connected to the Internet and able to receive messages from any node through the Internet or D2D connection.

\subsection{Architecture}\label{sec_hybrid_arch_desc}
The proposed network maximizes the potential of DTN and sparse Internet connectivity by utilizing the current or future connectivity of intermediary nodes to transmit information to destinations.
The hybrid network depends on the strength of the intermediary network for message delivery as the patients and their caregivers typically do not form a large enough density to facilitate efficient data delivery solely by opportunistic encounters.
Additionally, some of the nodes in the intermediary network have intermittent Internet connectivity that can be opportunistically leveraged to transmit medical data; resulting in increased delivery probability and decreased latency.

Figure~\ref{fig_opportunistic_path} is a depiction of a hybrid network where intermittent cloud connectivity is supplemented with delay tolerant opportunistic communication.
Time $t_0$ represents a discrete time and $t$ represents one or many instances 
of future times and/or durations such that $t= \big[t_0 ... t_{n-1}\big]$.
At time $t_0$, the patient's mobile device aggregated body sensor information 
from local sensors on the patients body along with survey data inputted by the patient.
In addition, the patient's mobile device attempted to communicate with the 
Clinical Staff, but there 
was neither Internet available nor was there a direct line-of-sight connection 
between them.
For the rest of the example, mobile devices are associated with their
respective owners, and will be referred to by their owners name, such as \textit{Patient}.
Also at time $t_0$, the Patient was in D2D range of Caregiver and forwarded their 
encrypted aggregated medical information intended for the Medical Entity, to Caregiver.
During $t_0<t\leq t_1$, the Caregiver encountered a member of the intermediary network and forwarded the Patient's 
encrypted medical information. 
Since no network participant encountered the Doctor while $t \leq t_1$ and 
no participating user carrying the Patient's message had an Internet 
connection, the data path to deliver the Patient's data to the Clinical Staff has 
not been established.

At a new time period, $t_2 = t > t_1$, all users were mobile resulting in 
new encounters in the network; the Caregiver to the medical entity, and a member of the intermediary network to a Clinical Staff. 
The Patient lost their D2D connection to their Caregiver and the D2D connection 
between the Caregiver and the intermediary node is no longer available.
In addition, the Medical Entity still has an Internet connection and the Clinical Staff 
has gained Internet access.
A new D2D connection is formed between an intermediary node and the Clinical Staff, who has an Internet connection.
Another D2D connection is also formed between the Caregiver and Medical Entity.
Hence, the Medical Entity is able to receive the Patient's data from the Internet or D2D; whichever occurs first.

\section{Modeling Rural Remote Patient Monitoring}\label{section:real_world}
In order to understand the viability and feasibility of the network, a simple mathematical model is introduced that depicts some real-world characteristics of the RRPM problem, including the high intermediate node to source-node ratio, limited number of data producers, prevalent presence of fully mobile caregivers and limited number of POIs.

\subsection{Model description}\label{model_desc}
Consider a network in which $N$ represents a set of nodes that are randomly distributed in a square grid divided into $M \times M$ cells, where an individual node is represented by $n \in N$. 
Let $\{A,S,C,I,P,D\} \subseteq N$
, where subsets are defined as: $A$ - patients, $S$ - clinical staff, $C$ - caregivers, $I$ - intermediary nodes, and $D$ - destinations, such that: 

\begin{equation}
    |D| \leq |S| \leq |A| \leq |C|  << |I| 
\end{equation}\label{equation:nodes}

\noindent Let the set $\{E,U\}\subseteq I$ represent employed and unemployed intermediary nodes respectively.
Furthermore, assume each node $n \in N$ has a connectivity parameter, $\omega$, that defines its current connectivity status where:

\begin{subnumcases}{\omega(n)=}
    \omega_{1} & \text{Internet available}\label{equation:connected}\\
    \omega_{2} & \text{D2D available}\label{equation:disconnected} 
           
\end{subnumcases}

\noindent 
In addition, assume that for the set $D$ that $\omega = \omega_{1}$, for set $A$ that $\omega = \omega_{2}$, and for $\{I,S\}$ there exists a probability, $r$, for which $\omega = \omega_{1}$ and \( 1 - r \) for which $\omega = \omega_{2}$.
The probability, $r$, of a node $n_r \in \{I,S\}$ having Internet connectivity at a certain time is determined by the rural community's broadband access rate.

Let an individual patient be represented by $a \in A$ and the set of messages for an individual patient be:

\begin{equation}
    m_{a,j}(t_{0}) \in m_a \mid 0 < j \leq |m_a|, t_{0} = t
\end{equation}

\noindent where message number $j$ is generated at time $t_0$ and is transmitted to $D$ at a time $t \leq t_{f}$.
For each message $m \in M$, there exists a set of nodes, $N_{m}$, that have a copy of message $m$ and a set of nodes $L_{m} = n-N_{m}$ that do not have the message.
At each distinct time, $t$ = [$t_{0}, t_{1},..., t_{f}$], encounters occur between nodes and through those encounters, messages in $M$ are transmitted.
Once a node in $N_{m}$ encounters a node, $l$, in $L_{m}$, the corresponding message, $m$, is transmitted and $l$ becomes a member of set $N_{m}$.

After $l$ obtains the message and is added to the node set, if $\omega(l)$ = $\omega_{2}$ and $l \notin D$, then nothing else happens.
However, if $\omega(l)$ = $\omega_{1}$ or if $l \in D$, the message is considered to be delivered and the difference between the start time and the time of which delivery occurs, $z_{m} = t - t_{0_m}$, is the delivery latency for the message, $m$.
At each consecutive time step, more encounters occur.
Finally, when $t = t_{f}$, the delivery probability can be calculated as the number of messages in $M$ transmitted to $D$.
\begin{equation}
    p_{\operatorname{delivery}} = \dfrac{|M_D|}{|M|}
\end{equation}
Additionally, the upper-bound delivery latency for all delivered messages is defined as the message with the largest $z$ or:
\begin{equation}
    z_{\max} = \max(z_m)
\end{equation}

\subsection{Mobility and transmission}\label{mobil}
The mobility of nodes in the network is described by $x$ discrete time Markov chains (DTMCs) with a finite number of states. 
When DTMCs are used with heterogeneous contact rates, they have been shown to approximate realistic mobility for DTN scenarios and scales well with network size~\cite{picu2012analysis}.
For simplicity, the following states are used: home, work, and POI. 
Individual home and work locations are assigned to each node and POIs can be randomly selected from the set, $P$, of POIs during each transition.

The subset \{D,P\} are considered stationary nodes and do not have a transition matrix associated with them. Each mobile node in subset $\{A,S,C,I\}$ has a unique transition matrix for each time period, $T(k) = \{t_{i}, t_{i+1}, ..., t_{\gamma}\} \mid 0 < k \leq x$. Where, each period $T(k)$ starts at $t_{i}$ and consists of $t_{\gamma} - t_{i}$ consecutive time steps.
For example, employed nodes such as $e \in E$ and $ s \in S$ are preferentially attached to and are stationary at work locations, which consists of POIs in the grid during the work period (e.g. 9:30am - 4:30pm).
%
%
Hence, $E$ and $S$ nodes are mobile between home and work.
The model assumes that contact occurs when two nodes with the same radio are within transmission range of each other where, the transmission range is assumed to be circular.
Messages are also assumed to be small enough to be successfully transmitted within each encounter and uniformly sized.

\begin{figure}[htbp]
	\centering
	\includegraphics[width=0.45\textwidth, height=2.715in]{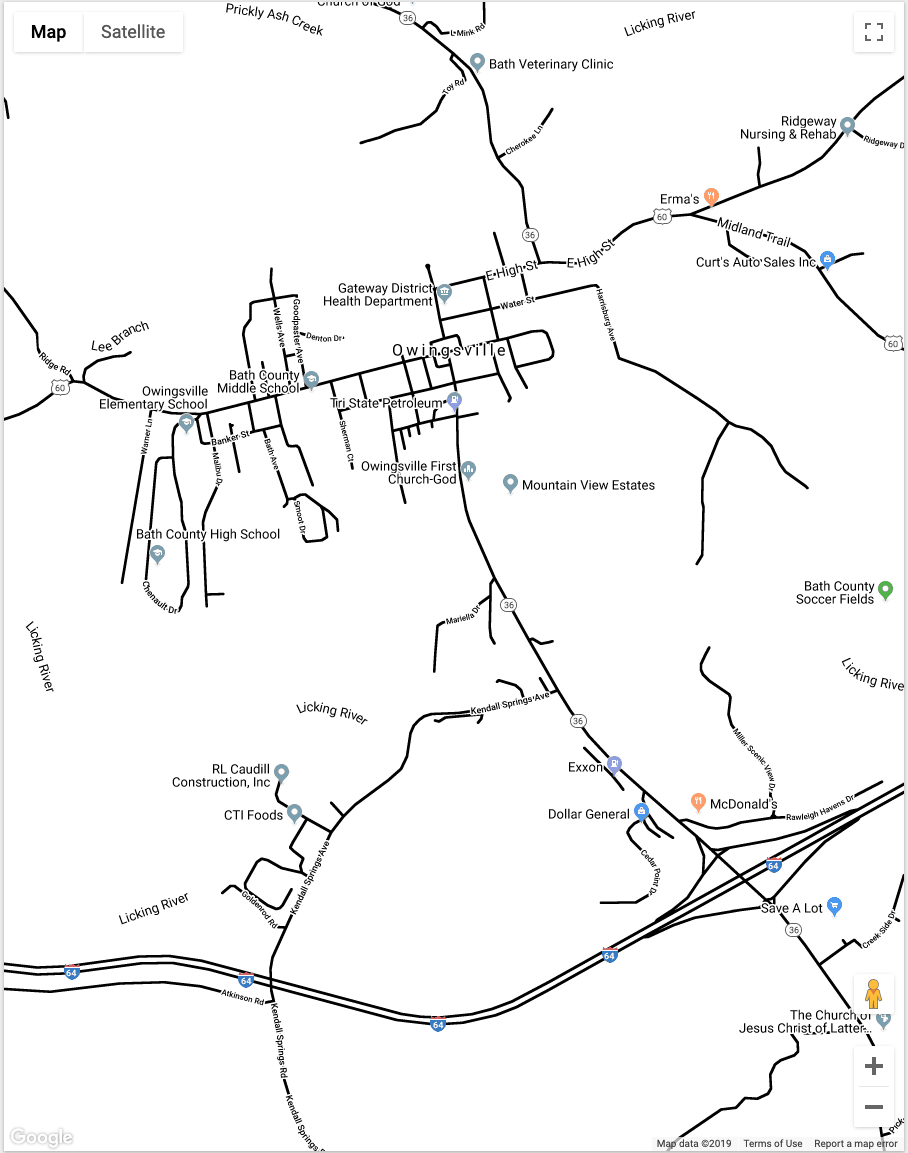}
	\caption{Map of Owingsville KY/Appalachian Kentucky.}
	\label{fig_owingsville_map}
\end{figure}

\section{Numerical Evaluation}\label{section:evaluation}
In evaluating the feasibility and viability of the model, data was obtained from the Federal Communications Commission and the US Census Bureau regarding Owingsville, KY (Bath County)\cite{census_quickfacts}.
Owingsville, KY was chosen because it is a rural city with health and connectivity issues while having a moderate technology adoption rate.
A map of Owingsville, KY can be found in Figure~\ref{fig_owingsville_map}.

\subsection{Medical Problem Domain}\label{subsection:medical_problem}
Distress monitoring for cancer patients was used as a domain example of RRPM due to its delay tolerant nature (messages are valid for over 24 hours) and the prevalence of cancer and distress within eastern Kentucky. 
Additionally, distress monitoring data from patient-to-provider and vice-versa are not considered emergency related or time-critical~\cite{services2017,max2019opportunistic}.
Likewise DTNs can reinforce or replace local infrastructure 
during emergencies and natural disasters\cite{syed2012study}.  
Overall, health data transmission such as distress screening, asynchronous 
triage, and dispatch of care are enabled by hybrid networks. 
The end result can lead to improved early recognition and appropriate response to 
disease, averting unnecessary care and avoidable adverse events for 
patients in challenging settings.

\subsection{Rural Node Mobility}\label{subsection:state_transitions}
As stated in Section~\ref{mobil}, maintaining heterogeneous contact rates between nodes 
in combination with DTMCs will result in realistic mobility~\cite{picu2012analysis}.
Contact rates in this work are based on a 2017 IPUMS ATUS sample of non-metropolitan households in the US, 303 routine activities were obtained, along with corresponding start and stop times, and classified into three states: Home, Work and POI~\cite{hofferth_flood_sobek}.
In addition, information from IPUMS ATUS was obtained for the number of individuals in each state for 30 minute intervals, and four (4) periods were defined based on the number of people in each state.
The four (4) periods defined can be found in the ``Time (Period)'' column in Table~\ref{transition_table}. 
Consequently, the transition matrix was estimated for each period by obtaining the transition matrix for each individual, and aggregating it over each period.
The resulting matrices and periods are given in Table~\ref{transition_table}.

\begin{table}[htbp] 
\centering
\caption{Transition Matrices Derived from ATUS Data.
}
\label{transition_table}
\resizebox{0.48\textwidth}{!}{%
\begin{tabular}{ccc}
\toprule[1.5pt]
Time (Period) & \multicolumn{2}{c}{ Initial Probability Vector and Transition Matrices ($Home$, $Work$, $POI$)}\\
\\
\midrule
 & \multicolumn{1}{c}{Node Classification: \{$C, U, A$\}} & \multicolumn{1}{c}{Node Classification: \{$E, S$\}}\\

\\
19:00 - 06:30 (1) & (0.85, 0, 0.015)\ $
\begin{pmatrix}
0.94 & 0 & 0.064\\
0 & 1 & 0\\
0.37 & 0 & 0.63\\
\end{pmatrix}
$ & (0.70, 0.079, 0.22)\ $
\begin{pmatrix}
0.85 & 0.019 & 0.13\\
0.14 & 0.81 & 0.043\\
0.39 & 0.32 & 0.58\\
\end{pmatrix}
$\\
\\
\hline
\\
\\
06:30 - 09:30 (2) & (0.93, 0, 0.070)\ $
\begin{pmatrix}
0.97 & 0 & 0.032\\
0 & 1 & 0\\
0.59 & 0 & 0.41\\
\end{pmatrix}
$ & (0.71, 0.16, 0.13) $
\begin{pmatrix}
0.86 & 0.079 & 0.061\\
0.17 & 0.61& 0.21\\
0.51 & 0.18 & 0.31\\
\end{pmatrix}
$ \\
\\
\hline
\\
\\
09:30-16:30 (3) & (0.76, 0, 0.24) $
\begin{pmatrix}
0.89 & 0 & 0.11\\
0 & 1 & 0\\
0.36 & 0 & 0.64\\
\end{pmatrix}
$ & (0.50, 0.33, 0.13) $
\begin{pmatrix}
0.80 & 0.083 & 0.12\\
0.063 & 0.90 & 0.037\\
0.30 & 0.057 & 0.64\\
\end{pmatrix}
$\\
\\
\hline
\\
\\
16:30-19:00 (4) & (0.77, 0, 0.23) $
\begin{pmatrix}
0.91 & 0 & 0.086\\
0 & 1 & 0\\
0.30 & 0 & 0.70\\
\end{pmatrix}
$ & (0.48, 0.20, 0.32) $
\begin{pmatrix}
0.80 & 0.027 & 0.17 \\
0.042 & 0.88 & 0.78 \\
0.28 & 0.058 & 0.66 \\
\end{pmatrix}
$\\
\\
\bottomrule[1.5pt]
\end{tabular}}
\end{table}

\subsection{Simulation setup}
To understand how the proposed hybrid network described in 
Section~\ref{model_desc} 
could be used for RRPM, distress information (described in Section~\ref{subsection:medical_problem}) from cancer patients (source 
nodes) to their respective healthcare providers (destination nodes) was used 
and a simulation environment was created in Python\footnote{Code available at \url{https://github.com/netreconlab/ICNC20}}.
Simulations were conducted to determine the effect of the number of patients 
and the number of intermediary nodes participating in the network on the 
delivery rate and delivery latency of messages.
The evaluation conducted provided the upper and lower bounds of the delivery 
rates and delivery latency respectively. 
To do this, we compared the effect of the aforementioned parameters on the 
delivery latency and delivery rates of three different network configurations.

The first was a DTN, similar to our previous work, \cite{maxUtilizing2019}, in which messages were transmitted to 
the destination solely through D2D communications via Bluetooth 5. 
The second was the proposed Hybrid Network, where, messages were transmitted 
both through D2D communications and the Internet connectivity of intermediary 
nodes.
The third was a user provided network (hereafter referred to as UPN 
configuration) that solely relied on the source node encountering a node with 
Internet connectivity or directly transmitting the message to the destination. 
A total of 100 seeds was used to randomize the simulation. 

Additionally, one message was generated per patient at the beginning of the 
simulation and each simulation began at randomly assigned period to better 
understand how distress monitoring information will propagate through the 
network.
To determine the upper/lower bounds of the delivery rates and delivery latency 
respectively, an ideal version of the Epidemic routing protocol was used that 
did not limit the abilities to broadcast or send information (i.e. buffer 
limitations) for the DTN and Hybrid scenarios.
Also, the strength of each node's radio signal was varied, with the 68\% of 
the nodes having a transmission range between 400ft and 800ft based on 
Bluetooth 5's theoretical range.
For the DTN, hybrid, and UPN configurations, patients were mobile, were the 
only generators of data as mentioned in Section~\ref{subsubsection:patient}, 
and never had Internet connectivity.
The number of cancer patients was varied from 2, which is the estimated number 
of lung cancer patients in Owingsville to 10, which is the estimated number 
of all cancer patients in Owingsville~\cite{state_profiles}.
Furthermore, the ratio of the population involved in the intermediary network 
was varied from 10\% to 100\% of the adults in the city.
%

In the DTN case, $\omega(n) = \omega_1$, for all nodes $n \in N$, where $N$ is 
a fraction of the number of people in the city's population from 10\% to 100\% 
of the population.
While the UPN case had $\omega(n) = \omega_2$, for all nodes $n \in N$, where
$N$ is a fraction of the number of people in the city's population from 2\% to 
20\% of the population.
Here $N$ represents the estimated number of people that have cellular 
connectivity, given that about 20\% percent of the population have cellular 
connectivity~\cite{pew_research_center_2019}.
The mobility patterns were estimated using transition matrices as describes 
in Section~\ref{subsection:state_transitions}.
A total of 25 employment locations/points of interest were randomly assigned at 
the start of each simulation.
Table~\ref{sim_parameters} describes the rest of the parameters used in the 
simulation along with the sources for their values where applicable.

\begin{table}[htbp]
\centering
\caption{Parameters used in Simulation}
\label{sim_parameters}
\resizebox{0.48\textwidth}{!}{%
\begin{tabular}{|l|c|c|}
\hline
\multicolumn{1}{|l|}{\textbf{Parameter}} & \multicolumn{1}{c|}{\textbf{Value}} & \multicolumn{1}{c|}{\textbf{Source}} \\ \hline
Simulation seeds & 0:1:99 &  --\\ \hline
Simulation duration & 24 hours & --\\ \hline
Adult Population of Owingsville & 400 & \cite{census_quickfacts}  \\ \hline
Area of Owingsville & 2.409 sqmi & \cite{census_quickfacts}  \\ \hline
Number of Cells & $820*820 = 672,400$ & \cite{census_quickfacts}  \\ \hline
Cell size & 10 ft $\times$ 10ft &  \cite{census_quickfacts} \\ \hline
Number of patients (\( |A| \)) & 2:2:10 &  \cite{state_profiles} \\ \hline 
Number of caregivers (\( |C| \)) & 2:2:10  & \cite{berry2016supporting} \\ \hline
Number of destinations (\( |D|\)) & 1  & \cite{berry2016supporting} \\ \hline
Ratio of population involved in intermediary network (\( I \)) & 0.1:0.1:1  & -- \\   \hline
Ratio of Internet connected intermediary nodes   & 0.2  & \cite{pew_research_center_2019} \\ \hline
Number of POIs (\( |P| \)) & 25  & Map (Figure~\ref{fig_owingsville_map}) \\ \hline
Number of Clinical Staff (\( |S| \)) & \( \leq 2 \)  & \cite{regulations_2014} \\ \hline
Periods & 1 to 4  & \cite{hofferth_flood_sobek} \\ \hline
Data generation rate & 1 message per 24 hours  & Markey Cancer Center \\ \hline
Ratio of employed nodes & 0.935  & \cite{labor_statistics} \\ \hline
Transmission range (based on Bluetooth 5) & \( \mu = 60,\ \sigma^2 = 20 \)  &  Bluetooth 5 SIG\\ \hline 
\end{tabular}}
\end{table}

\subsection{Results}\label{subsection:results}
\begin{figure}[htbp]
\centering
    
   \subfloat[Mean Delivery]{\includegraphics[width=0.24\textwidth,height=1.45in]{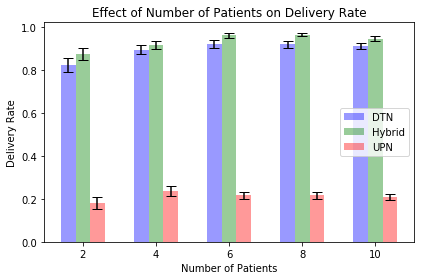}\label{Delivery_pats}}
    \subfloat[Mean Delivery Latency*]{\includegraphics[width=0.24\textwidth,height=1.45in]{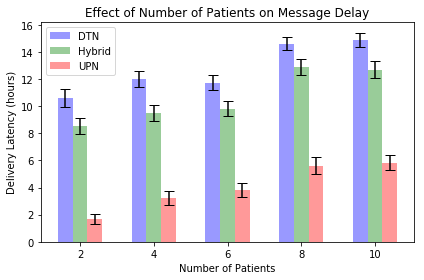}\label{Delay_pats}}
    
    \subfloat[Mean Delivery]{\includegraphics[width=0.24\textwidth,height=1.45in]{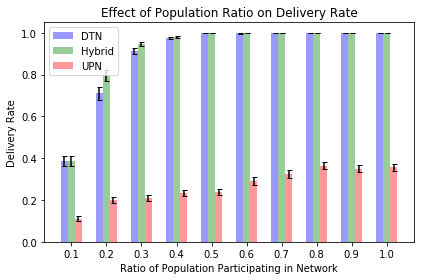}\label{Delivery_ratio}}
    \subfloat[Mean Delivery Latency*]{\includegraphics[width=0.24\textwidth,height=1.45in]{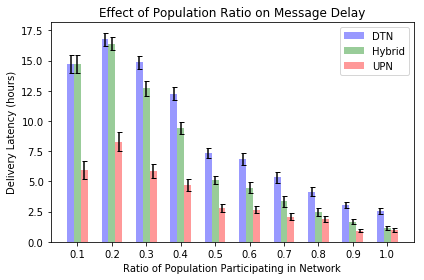}\label{Delay_ratio}}
\caption{(a) and (b) vary the number of patients in the network with .30 of the population participating in the network. (c) and (d) vary population participation in the network with 10 patients generating one message each with a time to live of 24 hours. In the Hybrid and UPN cases, .2 of the participating population have Internet connectivity. Error bars represent SEM. * shows the delay associated with delivered messages.}
\label{results}
\end{figure}

Figures~\ref{Delivery_pats} and \ref{Delay_pats}, give insight on the scalability of each network configuration when 30\% of the population participates in an intermediary network and 6\% of the population (20\% of the intermediary network) have cellular connectivity.
Regardless of the number of patients in each simulation, the hybrid network produces the highest delivery rate and the UPN configuration produces the lowest delivery rates among the different configurations.
Additionally irrespective of the number of patients in each simulation, the UPN configuration produces the lowest delivery latency and the DTN
configuration produces the highest delivery latency.
While delivery rate of the DTN configuration is 4.2\% less than the Hybrid configuration on average, the difference in delivery latency is 17.7\%.
Thus indicating that among the 3 configurations explored, the hybrid network is best equipped to handle higher priority messages.
Overall, increasing the number of patients has a positive effect on the DTN and hybrid network performance with a slight increase in delivery rate.
However, the delivery rate in the UPN configuration does not seem to be affected as the number of patients increases.

The effects of varying the number of intermediary nodes in the network on the delivery rate, delivery latency, are seen in Figures \ref{Delivery_ratio} and \ref{Delay_ratio}.
As more intermediary nodes are added to the network, the delivery rates of all configurations increase. 
However, the delivery rates of both the DTN and hybrid configurations increase logarithmically and eventually approach 100\% delivery when 50\% of the population participates in the intermediary network.
Similarly, the UPN configuration experiences an overall increase, howbeit, it never reaches 40\% delivery. 
Also, the DTN configuration consistently has the highest delivery latency, however, as the number of participants in the intermediary network increases, the delivery latency in the hybrid configuration rapidly reduces.
As a result, one sees that the hybrid network moves from nearly having the highest delivery latency (20\% participation) to nearly having the lowest delivery latency (100\% participation).

\section{Discussion and Future Work}\label{section:discussion}
Upon evaluation, the proposed hybrid network consistently shows significantly higher delivery rates at lower or comparable delivery latencies.
Based on the results in Section~\ref{subsection:results}, one can infer that 
the hybrid network may be most suitable for certain domain applications 
requiring a lower delivery latency. 
For example, a RRPM application that requires a delivery 
rate of .95 and a delivery latency with an upper bound of 3 hours may be 
most suitable for an hybrid network with 90\% of the population participating.
Conversely, an RRPM application that requires a delivery rate of .95 and 
a delivery latency with an upper bound of 2 minutes will require Internet 
connectivity for all devices and systems in the network.

While the proposed solution does provide an innovative means of transmitting 
patient data, it is not without limitations.
One such limitation is the need for population participation to increase the amount 
of intermediary nodes.
As a result of this limitation, the authors hope to explore a means of 
modeling node incentivization that represents human social behaviour.
In addition, the benefits of the hybrid model occur only in areas where there 
is sporadic Internet connectivity and assumes that healthcare providers have 
full connectivity.
If Internet connectivity is never available, the network will fall back to the DTN 
scenario.

Another limitation of this paper is it does not include the effects of privacy, security, ethics, and data use 
agreements as they are beyond the scope of this paper.
However, security measures can be implemented as described in other
work~\cite{baker2017vivo, max2019opportunistic}.
Lastly, the evaluations in this work does not consider different routing protocols to use for the DTN and Hybrid scenarios as the intent was to get an upper/lower bounds for delivery rate and delivery latency respectively. 
In order to optimize the network towards the upper/lower bounds in our findings, 
exploring well-known DTN protocols in real-world channel conditions will be a part 
of future studies.
The aforementioned future work will culminate into an applied evaluation and deployment of a RRPM system for lung cancer patients in Appalachian Kentucky. 

\section{Conclusion}\label{section:conclussion}
The authors propose a novel architecture that supplements intermittent 
Internet coverage by transmitting patients' health data opportunistically 
until it reaches healthcare providers. 
The simulation results, using real-world data from Owingsville, KY, a small 
rural Appalachian city, have demonstrated that the proposed model is feasible 
and can provide a timely and reliable communication to remotely link rural patients with 
their providers; resulting in better quality of care.

\bibliographystyle{IEEEtran}
\bibliography{TESCA}

\end{document}